\documentclass[english]{article}
\usepackage[T1]{fontenc}
\usepackage[latin9]{inputenc}
\usepackage{geometry}
\usepackage{color}
\usepackage{graphicx}
\usepackage{verbatim}
\usepackage{amssymb}
\usepackage{epstopdf}
\usepackage{amsmath}
\usepackage{mathtools}
\usepackage{caption}
\usepackage[normalem]{ulem}
\usepackage[makeroom]{cancel}
\usepackage{cite}
\usepackage{amsmath}
\usepackage[toc]{appendix}
\usepackage[colorlinks,citecolor=magenta,linkbordercolor={0 0 1}]{hyperref}
\newcommand{\comments}[1]{}   %%%%%%%%%%%%%%% comments

\makeatletter

\usepackage{indentfirst}
\usepackage{enumerate}

\makeatother

\begin{document}

\title{Dynamical typicality in classical lattice systems}

\author{Nicolas Nessi\\IFLP CONICET, Diagonal 113 y 64, La Plata, Buenos Aires, Argentina. \and Peter Reimann\\Faculty of Physics,
Bielefeld University,
33615 Bielefeld, Germany}

%\author{Nicol\'as Nessi$^1$\\
%$^1$ IFLP CONICET,\\
%Diagonal 113 y 64, La Plata, Buenos Aires, Argentina.
%}

\maketitle

\abstract{
Considering deterministic classical lattice systems with continuous
dynamical
variables, we show that, if the initial conditions are sampled according to a probability distribution in which the dynamical variables are statistically independent, the
%dynamical trajectory
temporal evolution
of \textit{any} macroscopic observable is approximately the same for the vast majority of
%the states in the sample.
the randomly sampled initial states.
Our proof relies on general concentration of measure results which provide tight bounds for the deviation from typical behavior in the case of large system sizes.
The only
%condition that we assume for
requirement on
the dynamics is that the influence
of a local perturbation in the initial state decays,
at any given instance of time,
sufficiently fast with distance.
%at any finite time.
%\PR{for all later time-point}.
Our results are relevant, in particular, to classical Hamiltonian systems on a lattice,
but also apply to a large class of dissipative dynamics.
We illustrate our general results with two analytical examples, a coupled map lattice and a Hamiltonian system of coupled rotors with long-range interactions.
}

%\newpage
%\tableofcontents

%\clearpage

\section{Introduction}
\label{sec:introduction}

In equilibrium statistical mechanics, the specification of a few macroscopic observables, such as energy, particle number, total momentum, etc., is enough to predict, with high accuracy, the value of other relevant quantities, such as various correlation functions. Far from equilibrium, outside the linear response regime, the situation seems much more complex, since, in principle, different initial states can lead to entirely different dynamical histories, regardless of their macroscopic features. However, it turns out that different initial states inside a properly defined set (drawn according to a certain probability distribution, sharing the value of some observable, etc.) tend to exhibit very similar \textit{dynamical} behavior. This \textit{dynamical typicality} phenomenon restores
a
certain symmetry between the equilibrium and out-of-equilibrium scenarios.

Dynamical typicality was first discovered for
%many-body
\textit{quantum} many body systems in
%Refs.~\cite{bartsch09_dyn_typ_first,reimann20_dyn_typ}.
Ref.~\cite{bartsch09_dyn_typ_first},
and was subsequently refined for instance in Ref.~\cite{reimann20_dyn_typ}
and further references therein.
%**agregar las referencias del paper de Reimann para engrosar la lista**.
%\PRR{What is the menaing of this?} \NNR{It says: Add references from Reimann's paper to enlarge reference list.}
In these systems, a strong form of typicality is valid: the vast majority of pure
initial
states with energies inside a microcanonical interval which at time $t=0$ have very similar
expectation
values of some set of observables $A_1, \ldots,A_K$ will exhibit very similar expectation values for \emph{any} other observable $O$ at any later time~\cite{reimann20_dyn_typ}, regardless of the nature of the observables $A_i$ defining the
%set
initial states
or the target observable $O$. This phenomenon is related to the geometric properties of high-dimensional spheres, which are isomorphic to the Hilbert spaces of many-body quantum
systems.
%For classical systems, according to very general arguments [cite new paper], dynamical typicality cannot hold for \textit{all} observables. Intuitively, microscopic observables in classical systems, such as the value of the momenta or position of one of the particles in a gas, show non-vanishing fluctuations in the thermodynamic limit, even in equilibrium. However, we can expect that suitably defined \textit{macroscopic} observables exhibit some kind of typicality. In fact, one of the authors have recently shown that for classical systems composed of a large number of binary state variables this was indeed the case~\cite{nessi25_dyn_typ_binary}.

It is interesting, then, to ask whether the same phenomenon occurs in \textit{classical} many-body systems and under which conditions. Some advance was made in Ref.~\cite{nessi25_dyn_typ_binary}, where the dynamical typicality phenomenon was addressed for systems consisting in large collections of binary
dynamical
variables. The purpose of this paper is to show how dynamical typicality
%arises for more general systems,
may arise for more general classical many-body systems,
in particular, for deterministic systems with continuous
dynamical
variables living on a lattice. We are able to show that, if the dynamics fulfills some reasonable conditions about the spatio-temporal spreading of perturbations, macroscopic observables indeed exhibit dynamical typicality. More precisely, the vast majority of initial states drawn according to an independent probability distribution of the dynamical variables show very similar dynamics for every macroscopic observable. The phenomenon of concentration of measure underpins our derivations. Concentration of measure is a phenomenon present in
%large
high
dimensional probabilistic spaces, by which functions which depend on many \textit{independent} random variables are essentially constant throughout
%the whole
nearly the entire space~\cite{talagrand96_new_look_independence,ledoux05_the_com_phenomenon,boucheron12_book_concentration}. In such cases, concentration of measure inequalities provide tight bounds for deviations from typical values for a large class of functions in the case of
%large
high dimensional system. The mathematical characterization of the set of functions which fall under the conditions of these general theorems provides us with a rigourous definition of what we call \textit{macroscopic observable}. Classical Hamiltonian lattice systems with interactions decaying sufficiently fast with distance are an important example in which our results can be applied. However, it is important to remark that a great variety of classical dynamical systems, such as coupled map lattices, also fall under the scope of our results.

The rest of the paper is organized as follows. In section~\ref{sec:general}, we first provide a minimal context about concentration of measure that will be helpful in the derivation of our main result, and then go on with the derivation itself, with special attention to the hypothesis involved. In section~\ref{sec:examples}, we develop two examples of applications of our results: a simple coupled map lattice and a Hamiltonian system of coupled rotors with long range interactions.
%For the later, we
For both examples, we also
provide numerical simulations that illustrate the results. Finally, in section~\ref{sec:conclusions}, we formulate our conclusions and outlook.

\section{General results}
\label{sec:general}

In this section we will present the main results of our paper. To motivate our
%derivation,
approach,
we will discuss first some general concentration of measure results, which constitute the core mathematical machinery behind our derivations.

\subsection{Concentration of measure}

Concentration of measure refers to a phenomenon present in
%large
high
dimensional probabilistic spaces (large collections of random variables). In such spaces, there are certain functions, to be characterized below, which are essentially constant throughout
%the whole
nearly the entire
space, provided that the variables are independently distributed. More precisely, for the overwhelming majority of points in the space, these functions take values which are very close to
%its expected, typical value.
their mean (or ``typical'') values.
In particular, the measure of the set of points which are far away from the
%expected value,
typical values
shrinks rapidly with the dimension of the space.
%The typical example of the class of functions that exhibit concentration
The most common examples of such functions
are \textit{sums} of random variables.
%For example,
For instance,
let $\left\{b_1,b_2,\dots,b_n\right\}$ be a collection of $n$ independently distributed binary random variables $b_i\in \{1,-1\}$, such that $P(b_i=1)=P(b_i=-1)=1/2$. Given that the variables are independent and that they have the same probability to be $1$ or $-1$, the expected value of their average, $s(b_1,b_2,\dots,b_n)
:=
\frac{1}{n}\sum_i b_i$, is zero, $E[s]=0$. The following \textit{concentration of measure inequality} bounds the probability of finding points which deviate from the expected value beyond a certain threshold $\epsilon>0$~\cite{talagrand96_new_look_independence}:
\begin{equation}
\label{eq:com_example}
\mathrm{Prob}\left( \left\vert \frac{1}{n}\sum_i b_i \right\vert>\epsilon \right)\leq 2\exp\left( -\frac{\epsilon^2 n}{2}\right).
\end{equation}
According to Eq.~(\ref{eq:com_example}), given any $\epsilon>0$, the probability that the average of the random variables is larger than $\epsilon$ is arbitrarily small for sufficiently large $n$. In other words, for sufficiently large $n$, the value of the average is sharply concentrated around $0$, its expected value.

Another familiar example of concentration of measure is the central limit theorem, according to which the relative variance of the average of a collection of $n$ independent real random variables
%vanishes as $n^{-1}$ as $n$ increases.
decreases asymptotically as $n^{-1}$ for large $n$.

Intuitively, if the function under consideration receives
notable
contributions from a large number of uncorrelated random variables, it is very unlikely that such variables can cooperate to pull the value of the function far away from the ensemble mean.

Although concentration of \textit{sums} of independent random variables, such as
in
the example considered above or the central limit theorem, was a fact known for a long time, major advances came when such a phenomenon could be proved for more general functions~\cite{talagrand96_new_look_independence,ledoux05_the_com_phenomenon}. As a prototypical result we can mention McDiarmid's inequality~\cite{boucheron12_book_concentration}, which we formulate in the following. Consider a collection of $n$ independently distributed real random variables $\vec{z}=(z_1,z_2,\dots,z_n)$, each one of them defined in a bounded interval $Z_i\subset \mathbb{R}$. Moreover, consider a function $g(\vec{z})$ which fulfills the so-called bounded differences condition
\begin{equation}
\label{eq:bounded_dif_a}
\vert \Delta g_i^{\mathrm{max}} \vert \coloneq \sup_{\vec{z},\hat{z}_i} \vert g(z_1,\dots,z_i,\dots,z_n)-g(z_1,\dots,\hat{z}_i,\dots,z_n) \vert \leq c_i,
\end{equation}
where the supremum is tacitly restricted to the above mentioned intervals.
This condition ensures that the maximum variation of the function when we modify the $i$-th argument is bounded by a finite constant $c_i$. Under such assumptions, the probability that the deviation of the value of $g$ from the mean
$E[g]$
exceeds
a
certain threshold $\epsilon>0$ is bounded according to McDiarmid's theorem:
\begin{equation}
\label{eq:mcdiarmid_a}
\mathrm{Prob}\left(  \vert g(\vec{z})-E[g] \vert\geq \epsilon  \right)\leq2\exp\left\{ -\frac{2\epsilon^2}{\sum_{i=1}^n\,c_i^2} \right\}.
\end{equation}

This result will play a pivotal role in our present work.
Besides its simplicity and generality, its most remarkable
feature is that the probability distribution of the
random variables $\vec z$ and the properties of the function
$g(\vec z)$ only enter via the constants $c_i$
in (\ref{eq:bounded_dif_a}).
%The simplest example of application of McDiarmid's inequality
Once again, the simplest example of such a function $g(\vec z)$
is the average $s(\vec{z}):=\frac{1}{n}\sum_{i=1}^{n}z_i$.
%of the variables.
In
%such
this
case, $c_i=\vert Z_i \vert/n$ and,
%clearly,
therefore,
$\sum_i\,c_i^2\sim n^{-1}$
under the tacit extra assumption that there exists an $n$-independent upper bound for all $\vert Z_i \vert$.
%which,
According to Eq.~(\ref{eq:mcdiarmid_a}) this implies
that the average $s(\vec{z})$ is sharply concentrated around its mean $E[s]$ for large $n$.

Another relevant result for our discussion is the Efron-Stein inequality,
which ensures that, under the same conditions as
in
McDiarmid's theorem, the variance of the function $g$ is bounded by~\cite{boucheron12_book_concentration}
\begin{equation}
\label{eq:es}
    \mathrm{Var}[g]
    :=
    E[(g(\vec{z})-E[g])^2]\leq \frac{1}{4}\sum_{i=1}^n\,c_i^2.
\end{equation}
In the example of the average, this implies that the relative variance $\mathrm{Var}[s]/E[s]^2$ scales as $n^{-1}$, in accordance with the central limit theorem.
%\PRR{What if $E[s]=0$ ?} \NNR{In such cases I think that you have to study the behaviour of the
%variance, without using the square mean as normalization.}

%We are interested in applying
Our main objective is to apply
these powerful
concentration of measure
results to the
%study
dynamics
of many-body classical systems. Roughly speaking, we will identify the random variables $\vec{z}$ with the dynamical variables of a given many-body system living on a lattice and the function $g$ with what we will call a \textit{macroscopic observable}, a function of the dynamical variables that receives contributions from all of them but does not put too much weight on any one in particular. In the next section we will formalize this correspondence
%through a series of assumptions about
by means of a series of conditions imposed on
the dynamics, the observables and the initial conditions under consideration.

\subsection{Concentration of measure applied to the dynamics of many-body classical lattice systems}

We consider the dynamics of a deterministic classical system with continuous
dynamical
variables $x_i$, with $i=1,\dots,n$, where index $i$ labels sites on a given lattice. The state of the system is described by a vector $\vec{x}=(x_1,\dots,x_n)$ in the product space $\mathbb{X}=X_1\times\dots\times X_n$, where $X_i$ is a given, bounded interval on the real line.
For instance, the vectors $\vec{x}=(x_1,\dots,x_n)$ may amount to phase space points of a
Hamiltonian system, but also other types of dynamical systems, such as coupled map
%systems,
lattices,
are admitted (see section \ref{sec:examples}).
We will formally denote the dynamics of the system by a transformation $\Phi_t:\mathbb{X}\rightarrow \mathbb{X}$, that maps the initial conditions
%into
onto
the state at time $t$
\begin{equation}
\vec{x}(t)=\Phi_t (\vec{x}(0))
\label{5}
\end{equation}
with the usual properties $\Phi_t\circ \Phi_s=\Phi_{t+s}$ for all $t,s$ and $\Phi_0=1$.

Note that the time $t$ is admitted to be either continuous or discrete.
Furthermore, we will tacitly focus on times $t\geq 0$, though in many cases
an extension to $t<0$ is straightforward.
Finally, our present approach is
%largely
independent of whether the
considered system is energy-conserving or dissipative.
Likewise, other common system properties like chaos, mixing, or ergodicity
will not play any prominent role.
Instead, the key requirements will be that the system is
of large dimensionality, that the considered observables are
of macroscopic character, and that the dynamics
%%satisfies a certain
%exhibits a
%%certain insensitivity against
%%limited sensitivity against
%suitably limited response to
%sufficiently remote local perturbations, as specified in more detail below.}
is such that the influence
of a local perturbation in the initial state decays,
at any given later instance of time,
sufficiently fast with distance
(for more details see below).
%at any finite time.
%\PR{for all later time-point}.

Our
%proof
approach
relies on the general concentration of measure results
from the previous section
which provide tight bounds for the deviation from typical behavior in the case of large system sizes.
%The only condition that we assume for the dynamics is that the influence of a local perturbation in the initial state decays,
%\PR{at any given instance of time},
%sufficiently fast with distance.
%%at any finite time.
%%\PR{for all later time-point}.
%
%Moreover, we will assume
%A
Accordingly, yet another
key aspect of our present approach is the assumption
that the initial conditions
$\vec{x}(0)$ are randomly
drawn from a product probability measure, in such a way that $x_i(0)$ are statistically independent random variables
(on bounded intervals $X_i$, as already said).
% Although the dynamical variables $x_i$ do not need to be bounded, we require the distribution of initial conditions for each $x_i(0)$ to be supported in a bounded interval $X_i$ in the real line.
For the rest, the probability measure can still be chosen largely arbitrarily.
We also note that this choice of the initial conditions is quite different from the corresponding choice in the
context of dynamical typicality for quantum systems~\cite{bartsch09_dyn_typ_first,reimann20_dyn_typ}, see also
section
\ref{sec:introduction}.
%\PRR{Can we say more about this ensemble of initial states? For instance, do most of them have very similar energies?
%Are most of them (or at least the average over all of them) far from equilibrium?
%Otherwise there is in my opinion little difference between ordinary (non-dynamical) typicality and dynamical typicality.
%In fact, Fig. 1 suggest to me that we are not very far from equilibrium at $t=0$.} \NNR{I've tried to give
%more details about the initial conditions in the paragraph below Eq.~(\ref{eq:mcdiarmid}).}

We will investigate the dynamics of observables represented by state functions $g:\mathbb{X}\rightarrow \mathbb{R}$. In particular, we will be interested in what we will call \textit{macroscopic observables}.
%As mentioned in
As already announced at the end of
the previous section, intuitively, a macroscopic observable is associated with a function $g$ which
%depends on all the microscopic coordinates,
generally may depend on all the microscopic state variables $x_i$,
but does not put too much weight on any one of them. To formalize this
%description,
property,
%we will ask $g$ to have bounded derivatives,
our first requirement is that $g$ must have bounded derivatives, meaning that
\begin{equation}
\label{eq:finite_der}
    \left| \frac{\partial g(\vec{x})}{\partial x_i}\right| \leq L<\infty,\, \forall i
    \in\{1,...,n\}
    ,\forall \vec{x}\in \mathbb{X}
\end{equation}
for some $n$-independent constant $L$.
It is easy to check that this implies that $g$ satisfies
%the so-called bounded differences condition
a very similar kind of bounded difference condition as in Eq.~(\ref{eq:bounded_dif_a}), namely,
\begin{equation}
\label{eq:bounded_dif}
\vert \Delta g_i^{\mathrm{max}} \vert \coloneq \sup_{\vec{x},\hat{x}_i} \vert g(x_1,\dots,x_i,\dots,x_n)-g(x_1,\dots,\hat{x}_i,\dots,x_n) \vert \leq c_i .
\end{equation}
The reason is that
%since,
if Eq.~(\ref{eq:finite_der}) is verified, we can choose $c_i=L\,\vert X_i \vert$, where $\vert X_i \vert$ is the length of the interval. This property ensures that the modification of one argument has a bounded effect on the value of the function.
%Moreover, we will assume
Our second requirement concerning $g$ is
that the
%finite \PRR{why is ``finite'' important here?} \NNR{I think that what I tried to say is that the $c_i$s should
%not be all equal to zero because they are in the denominator of the exponent in McDiarmid's inequality.
%But now that you raise the point, I think it is a completely disposable clarification.}
constants $c_i$ scale with $n$ in such a way that
\begin{equation}
\label{eq:ci_limit}
    \lim_{n\to\infty}\sum_{i=1}^n\,c_i^2=0.
\end{equation}
This
%last
condition ensures that, in the thermodynamic limit
$n\to\infty$,
no microscopic variation can produce a macroscopic effect on the function.
%In short,
Altogether,
conditions
%Eqs.~(\ref{eq:finite_der}), (\ref{eq:bounded_dif}) and (\ref{eq:ci_limit})
(\ref{eq:finite_der})-(\ref{eq:ci_limit})
give a rigorous definition of macroscopic observables. A characterization based in terms of Lipschitz functions is also possible~\cite{nessi25_dyn_typ_binary}.
%\PR{From now on, we always take it for granted that $g$ is a macroscopic observable of this type.}

% The simplest example of a macroscopic observable is the mean $f=\frac{1}{n}\sum_{i=1}^{n}x_i$ of the variables. In such case, $c_i=\vert X_i \vert/n$ and,
%clearly,
%therefore,
%$\sum_i\,c_i^2=\mathcal{O}(n^{-1})$
%under the tacit extra assumtion that there exists an $n$-independent upper bound for all $\vert X_i \vert$.
%\PRR{I did not understand why it is forbidden to choose for instance  $|X_i|=i$, implying  $\sum_i\,c_i^2 \geq 1$.} \NNR{In general, we want condition Eq.~(\ref{eq:ci_limit}) to be verified in order to have sharp concentration in the large $n$ limit. If $\sum_i\,c_i^2 \geq 1$ this is not possible.}

Given a macroscopic observable $g$, and since we are assuming that the initial conditions for the microscopic degrees of freedom are independently distributed, McDiarmid's inequality Eq.~(\ref{eq:mcdiarmid_a}) guarantees that at
time
$t=0$ the values of the observable are distributed around the expected value
%$E[f]$
$E[g]$
according to
\begin{equation}
\label{eq:mcdiarmid}
\mathrm{Prob}\left(  \vert g(\vec{x}(0))-E[g] \vert\geq \epsilon  \right)\leq2\exp\left\{ -\frac{2\epsilon^2}{\sum_{i=1}^n\,c_i^2} \right\}.
\end{equation}
Moreover, according to Eq.~(\ref{eq:ci_limit}), concentration becomes infinitely sharp for increasing $n$. In other words, the mean is the typical value of the observable.

%Notice that,
The inequality (\ref{eq:mcdiarmid}) implies that,
for any initial product probability distribution, all macroscopic observables exhibit sharp concentration around their mean as $n$ becomes large. Consequently, given a set of macroscopic observables, each
such product probability measure
%over microscopic initial conditions
can be labeled with the typical values of the set of macroscopic observables in this limit. Such typical values depend on the details of the probability distribution. For example, in a system with Hamiltonian dynamics, and assuming that the energy is a macroscopic observable (see section~\ref{sec:examples} for an example), every product measure for the microscopic variables will be associated with a value for the energy. This does not imply, however, that such product measure will be related to the microcanonical measure, since other macroscopic observables may concentrate around typical values which are different from the
%equilibrium
microcanonical
ones. In other words, changing the details of the probability distribution, we can define infinitely many initial ensembles which are, in general, out of equilibrium.

Incidentally, these findings for $t=0$ are closely related to ``ordinary'' (non-dynamical) typicality phenomena in
classical many-body systems, as explored for instance in Refs.~\cite{nessi25_dyn_typ_binary,rei25},
%and references therein,
the main difference being the details of how the initial conditions are chosen.
In spite of these differences, the findings in those previous works are in agreement with our above results in the case $t=0$.

%For classical systems, according to very general arguments [cite new paper, {\bf arxive}], dynamical typicality cannot hold for \textit{all} observables. Intuitively, %microscopic observables in classical systems, such as the value of the momenta or position of one of the particles in a gas, show non-vanishing fluctuations in the %thermodynamic limit, even in equilibrium. However, we can expect that suitably defined \textit{macroscopic} observables exhibit some kind of typicality. In fact, one of the authors have recently shown that for classical systems composed of a large number of binary state variables this was indeed the case~\cite{nessi25_dyn_typ_binary}.

The core subject of this paper is to elucidate under which conditions
the above reported concentration effect
%for $t=0$ persists for finite times.
at the initial time $t=0$ also persist for later times $t>0$.
In order to address this problem, we will first define the time evolved observable
\begin{equation}
g_t(\vec{x}(0))\coloneq\, g(\Phi_t (\vec{x}(0)))= g(\vec{x}(t)).
\label{6}
\end{equation}
Given that we want to exploit the statistical independence of the initial conditions,
the key point in the definition (\ref{6}) is that we
consider $g_t$ to be a function of $\vec{x}(0)$.
Then, our problem reduces to investigate which conditions ensure that $g_t$ is a macroscopic
observable in the sense of (\ref{eq:finite_der})-(\ref{eq:ci_limit})
%for $t>0$, i.e., if conditions
%Eqs.~(\ref{eq:finite_der}), (\ref{eq:bounded_dif}) and (\ref{eq:ci_limit}) are fulfilled at
%finite times during the evolution of the system.
for any given $t>0$.
Consequently, in order to verify the bounded differences condition, Eq.~(\ref{eq:bounded_dif}), we need to investigate the variation of $g_t$ when one of the microscopic coordinates of the initial condition is changed
\begin{equation}
\Delta g^{i}_t(\vec{x}(0),\Delta x_i(0))\coloneq g_t(x_1(0),\dots,x_i(0),\dots,x_n(0))-g_t(x_1(0),\dots,x_i(0)+\Delta x_i(0),\dots,x_n(0)).
\label{12}
\end{equation}
Assuming that
%the dynamics is
all pertinent functions are
differentiable,
%we have
we can infer from (\ref{6}) that
\begin{equation}
\frac{\partial g_t(\vec{x}(0))}{\partial x_i(0)}=\frac{\partial g(\vec{x}(t))}{\partial x_i(0)}=\sum^{n}_{j=1}\left.\frac{\partial g(\vec{z})}{\partial z_j}\right|_{\vec{z}=\vec{x}(t)}\,\frac{\partial x_j(t)}{\partial x_i(0)}.
\end{equation}
Together with Eq.~(\ref{12}) and the fundamental theorem of calculus, we get
\begin{equation}
\label{eq:deltaft}
\Delta g^{i}_t(\vec{x}(0),\Delta x_i(0))=\int^{\Delta x_i(0)}_{0}\,\sum^{n}_{j=1}\left.\frac{\partial g(\vec{z})}{\partial z_j}\right|_{\vec{z}=\vec{x}(t)}\,\frac{\partial x_j(t)}{\partial x_i(0)}\,dx_i(0).
\end{equation}
We see that the factors $\frac{\partial x_j(t)}{\partial x_i(0)}$ have naturally appeared. They are the response functions that measure how a perturbation in one coordinate at $t=0$ impacts on another coordinate at time $t>0$. Now, we will bound Eq.~(\ref{eq:deltaft}). Using the triangle inequality,
\begin{equation}
\label{eq:bound_deltag}
\vert \Delta g^{i}_t(\vec{x}(0),\Delta x_i(0))\vert
\leq
\left|
\int^{\Delta x_i(0)}_{0}\,
\sum^{n}_{j=1}\left|\frac{\partial g(\vec{z})}{\partial z_j}\right|_{\vec{z}=\vec{x}(t)}\,\left|\frac{\partial x_j(t)}{\partial x_i(0)}\right|\,dx_i(0)
\right|
,
\end{equation}
where the outermost absolute signs on the right-hand side
are only needed in cases where the upper integration limit
$\Delta x_i(0)$ is negative.

First, we note that the factor $\left|\frac{\partial g(\vec{z})}{\partial z_j}\right|_{\vec{z}=\vec{x}(t)}$
 inside the sum is bounded according to Eq.~(\ref{eq:finite_der}).
%Now,
Second,
we define
\begin{equation}
D_{ij}(t)\coloneq \left|\frac{\partial x_j(t)}{\partial x_i(0)}\right|.
\label{15n}
\end{equation}
%we get
%\begin{equation}
%\vert \Delta f^{i}_t(\vec{x},\Delta x_i(0)) \vert < L\sum^{n}_{j=1}D_{ij}(\Delta x_i(0);t).
%\end{equation}
To proceed, we need to characterize these quantities.
It is reasonable to expect that, in general, for fixed $i$ and $j$, they will be increasing functions of time.
In fact, in chaotic systems
their growth is bounded by $e^{\lambda_{\mathrm{max}}t}$, where $\lambda_{\mathrm{max}}$ is the largest Lyapunov exponent. On the other hand,
%at fixed time
at any fixed time-point $t$
it is reasonable to expect that
they decrease with
%the
increasing
distance between $i$ and $j$, which is related to the fact that perturbations
should
travel at finite velocities. The decay rate depends on the range of the interactions in the system. The interplay between the temporal and spatial dependencies of these quantities gives rise to a light-cone effect in the spreading of correlations in lattice systems with short range interactions
or with long range interactions that decay fast enough~\cite{kastner14_spreading_long_range}.
Taking these considerations into account,
%our hypothesis
the condition that we impose
on the dynamics will be that the decay of $D_{ij}(t)$ with distance, at fixed time, is sufficiently fast to ensure that the sum $\sum^{n}_{j=1}D_{ij}(t)$ is convergent and that there exists a uniform bound on it, independent of $\vec{x}(0)$ and $n$, i.e.,
\begin{equation}
\label{eq:dyn_hyp}
\sum^{n}_{j=1}D_{ij}(t)<K_i(t),
\end{equation}
where $K_i(t)$ is an
%$\mathcal{O}(n^0)$
$n$-independent
function of time.

More precisely speaking, the property (\ref{eq:dyn_hyp}) represents our main precondition
on the dynamics which guarantees the emergence of dynamical typicality for macroscopic observables.
Moreover, it seems intuitively reasonable to hypothesise that this property will indeed be fulfilled
in many physically interesting cases.
For instance,
%It is important to note that Eq.~
condition (\ref{eq:dyn_hyp}) is fulfilled by any dynamical system for which
small finite initial differences between two points do not get mapped to macroscopic (extensive) differences at
%\textit{finite times}.
any later time-point $t>0$.
%\PRR{I did not understand this statement since (9) includes the absolute value, but (8) does not. Please explain in somewhat more detail.} \NNR{I've added an intermediate step below Eq.~(\ref{eq:deltaft}).}
%The hypothesis is then
%\PRR{Maybe the following sentence should be omitted here in view of the concerns of the
%referee if there are multiple attractors. Similar concerns might be raised or if the
%considered (Hamiltonian) system exhibits a phase transition in the
%thermodynamic limit. Instead we could discuss this issue in somewhat more
%detail in the conclusions.} \NNR{Ok.}
%\PR{In particular, condition Eq.~(\ref{eq:dyn_hyp}) is thus}
%expected to be verified in any dynamical many-body system with short range interactions or with long range interactions that decay fast enough.
%\PRR{The latter case seems not covered above (10).} \NNR{The sum in Eq.~(\ref{eq:dyn_hyp}) can be convergent even for systems with long range interactions, as in the example in the next section}.
%\PRR{I tried to clarify my (very minor) point by the little extra text above (11).}

%Then, if the hypothesis is fulfilled,
Taking for granted that condition Eq.~(\ref{eq:dyn_hyp}) is fulfilled, we will
now bound $\vert \Delta g^{i}_t(\vec{x},\Delta x_i(0)) \vert$ from
Eq.~(\ref{eq:bound_deltag}). In order to do that, first consider
that $|\Delta x_i(0)|<|X_i|$, where $\vert X_i \vert$ is the
length of the interval $X_i$
where the dynamical variable $x_i$ takes support
(see also above Eq.(\ref{5})).
Moreover, using Eqs.~(\ref{eq:finite_der}) and~(\ref{15n}), we find
%Taking for granted that condition Eq.~(\ref{eq:dyn_hyp}) is fulfilled, it follows from Eqs.~(\ref{eq:finite_der}), (\ref{eq:bound_deltag}) and~(\ref{15n}) that
%\PRR{I think that also (6) and (15) must be used. Moreover, one has to observed that
%$|\Delta x_i(0)|<|X_i|$ and thus one still has to remember the meaning of $X_i$
%and $|X_i|$, which may possibly not be takeen for granted! Please amend if you agree.}
%\PRR{please provide the equation numbers which are utilized in this conclusion}
\begin{equation}
\vert \Delta g^{i}_t(\vec{x},\Delta x_i(0)) \vert < c_i(t),
\label{17}
\end{equation}
with
\begin{equation}
\label{eq:c_dyn}
c_i(t)
:=
L\,K_i(t)\vert X_i \vert.
\end{equation}

%\NNR{I reformulated the arguments that followed, hopefully bringing more clarity. If not, we can continue working with it.}

Focusing on an arbitrary but fixed $t>0$, and observing that the right-hand side of
Eq.~(\ref{eq:c_dyn}) is independent of $\vec{x}$ and $\Delta x_i(0)$, we see that
Eq.~(\ref{17}) is tantamount to Eq.~(\ref{eq:mcdiarmid}), but now applying to the
observable $g_t(\vec{x}(0))$ from Eq.~(\ref{6}).
Therefore, our natural next goal is to verify
condition (\ref{eq:ci_limit}), but now for the quantities $c_i(t)$ instead of the $c_i$.
%\PR{Upon comparing Eqs. (11) and (17) with Eq. (7), our natural next goal is to verify
%condition (\ref{eq:ci_limit}), but now for the quantities $c_i(t)$ instead of $c_i$
%on the left-hand side, which will then allow us to infer an inequality analogous to (9).
%As a first step,
% it is thus central to understand how the $c_i(t)$ scale with $n$.}
%to our argument to understand the scaling properties of $c_i(t)$ with respect to $n$.}
%, in order to verify condition in Eq~(\ref{eq:ci_limit}).
%Given that $c_i=L\,\vert X_i \vert$,
%we find,
To this end, we can employ Eq.~(\ref{eq:c_dyn}) to conclude
\begin{equation}
\label{eq:scaling_analysis}
%\sum_i
\sum_{i=1}^n
\, c^2_i(t)=
%\sum_i
\sum_{i=1}^n
\,L^2\,K^2_i(t)\vert X_i \vert^2\leq K^2_{\mathrm{max}}(t)
%\sum_i
\sum_{i=1}^n
\,L^2\,\vert X_i \vert^2=K^2_{\mathrm{max}}(t)\,
%\sum_i
\sum_{i=1}^n
\,c^2_i,
\end{equation}
where $K^2_{\mathrm{max}}:=\sup_{i}\,K^2_i(t)$,
and where we exploited in the last step that $c_i=L\,\vert X_i \vert$,
as mentioned below Eq.~(\ref{eq:bounded_dif}).
Recalling
the fact that $K_i(t)$ has no $n$-dependence
(see text
%above
below
Eq.~(\ref{eq:dyn_hyp})), we caninfer from
Eqs.~(\ref{eq:ci_limit}) and (\ref{eq:scaling_analysis}) that,
% at any finite time $t$,
for any given $t>0$,
\begin{equation}
\label{eq:ci_limit_dyn}
\lim_{n\to \infty}
%\sum_i
\sum_{i=1}^n
\,c^2_i(t)=0.
\end{equation}
Upon comparing Eqs.~(\ref{17}) and (\ref{eq:ci_limit_dyn})
with (\ref{eq:bounded_dif}) and (\ref{eq:ci_limit}) we can conclude that
we achieved the goal announced below Eq.~(\ref{6}) to show
that $g_t$ is a macroscopic observable for any
given
%finite value
$t> 0$,
and that, consequently, its values will be sharply concentrated around the mean for large $n$ according to
\begin{equation}
\label{eq:mcdiarmid_dyn}
\mathrm{Prob}\left(  \vert g_t(\vec{x}(0))-E[g_t] \vert\geq \epsilon  \right)\leq2\exp\left\{ -\frac{2\epsilon^2}{\sum_{i=1}^n\,c_i(t)^2} \right\} .
\end{equation}
%which is only Eq.~(\ref{eq:mcdiarmid}) written for the specific case of the observable $g_t(\vec{x}(0))$.

Let us finally consider the opposite case that $n$ is possibly very large but \textit{fixed},
while $t$ is variable.
%We will now consider the case of \textit{fixed} $n$.
In such case, since the functions $D_{ij}(t)$ from (\ref{15n}) may often be expected to asymptotically
diverge for large $t$ (see also the discussion below Eq.~(\ref{15n})), and the same
behavior will be inherited by the functions $K_j(t)$ in Eq.~(\ref{eq:dyn_hyp}), there will be a competition between $K^2_{\mathrm{max}}(t)$ and
$\sum_{i=1}^n\,c^2_i$
%in
on the right-hand side of
Eq.~(\ref{eq:scaling_analysis}). Indeed, considering Eq.~(\ref{eq:mcdiarmid_dyn}),
given any $\epsilon>0$, for any fixed value of $n$, there will always exist a
sufficiently large value of time such that
$K^2_{\mathrm{max}}(t)\,\sum_{i=1}^n\,c^2_i\gg 2\epsilon^2$.
%In such case, the bound becomes completely uninformative.
In other words, for any given, possibly very large but finite value of $n$,
our bound Eq.~(\ref{eq:mcdiarmid_dyn}) ultimately becomes
uninformative for sufficiently large $t$.

Additionally, we note that under the conditions that we are working, the Efron-Stein inequality
(\ref{eq:es}) provides a bound for the variance of
$g_t$~\cite{boucheron12_book_concentration}
\begin{equation}
\label{eq:es_dyn}
    \mathrm{Var}[g_t]
    :=
    E[(g_t-E[g_t])^2]\leq \frac{1}{4}\sum_{i=1}^n\,c_i(t)^2.
\end{equation}

%It is important to note that, at fixed $n$, given that $K_i(t)$ is an increasing function of time, our bounds Eqs.~(\ref{eq:mcdiarmid}) and~(\ref{eq:es_dyn}) eventually become %uninformative for sufficiently long times.

%Although our findings apply to macroscopic observables in chaotic systems, there is no need to
%assume mixing or chaos to obtain our results. Moreover, it is important to
We also note that our derivation provides sufficient
%and
but
not necessary conditions for the emergence of dynamical typicality. In particular,
there might be
%systems which fall outside of our hypothesis
examples of observables and/or dynamical systems
which do not satisfy our present key conditions
(\ref{eq:finite_der})-(\ref{eq:ci_limit}) and (\ref{eq:dyn_hyp})
but still exhibit the dynamical typicality phenomenon.

In the following section we will analyze two concrete examples which illustrate
%this
these
general arguments.

\section{Examples}
\label{sec:examples}

\subsection{Coupled map
%lattice
lattices
in one dimension}

Coupled map lattices are particularly
simple dynamical models
%of real variables
with continuous (real-valued) dynamical variables,
living on a lattice,
and evolving in discrete time.
They are generalizations of cellular automata, for which
the dynamical
variables only take discrete values. They are used as models of spatially extended chaotic systems~\cite{katzav05_cml_diff_eqs}, and also emerge in the study of quantum field theories~\cite{beck95_chaotic_quantization_FT}. We will consider a simple, one-dimensional coupled map lattice whose dynamics is given by the following recurrence relation,
\begin{equation}
\label{eq:cml}
x_i(t+1)=(1-\gamma)f(x_i(t))+\frac{\gamma}{2}\left[ f(x_{i-1}(t))+f(x_{i+1}(t)) \right],
\end{equation}
where $x_i$, with $i=0,\dots,n$ are the dynamical variables, $0<\gamma\leq 1$ and $t$ is the discrete time index.
We assume periodic boundary conditions, $x_0=x_{n+1}$. For simplicity, we will
moreover
assume that the dynamical variables are bounded and take values inside the same finite interval $X$, $x_i\in X$ for all $i$. %Moreover,
Finally,
we will assume that the derivative of the local map function $f$ is bounded,
\begin{equation}
\label{eq:cond_local_map}
\left\vert \frac{d f}{dx} \right\vert\leq M,\;\forall x\in X,
\end{equation}
where $M$ is a positive, finite
% \PRR{why is ``finite'' important here?} \NNR{Because if $M$ is infinite
%the bound is meaningless.}
 constant. In order to verify our main
%hypothesis in Eq.~
condition
(\ref{eq:dyn_hyp}), we need to
%bound $\sum_{j}D_{ij}(t)$, where $D_{ij}(t)=\left\vert \frac{\partial x_j(t)}{\partial x_i(0)} \right\vert$.
evaluate the quantities in Eq.~(\ref{15n}).
%From the recurrence relation and given the bound over the local map function derivative, we have
Therefore, we employ Eqs.~(\ref{eq:cml}) and (\ref{eq:cond_local_map}) to conclude
\begin{equation}
\label{eq:bound_cml}
\left\vert \frac{\partial x_i(t+1)}{\partial x_j(t)} \right\vert\leq M\left[ (1-\gamma)\delta_{ij}+\frac{\gamma}{2}(\delta_{i-1,j}+\delta_{i+1,j}) \right].
\end{equation}
If we group the dynamical variables into one vector $\vec{x}=(x_1,x_2,\dots,x_3)$, we can define the one-step Jacobian matrices as $J(s)=\frac{\partial \vec{x}(s+1)}{\partial \vec{x}(s)}$, i.e., $[J(s)]_{ij}=\frac{\partial x_i(s+1)}{\partial x_j(s)}$. Using the chain rule we get,
\begin{equation}
\frac{\partial \vec{x}(t)}{\partial \vec{x}(0)}=J(t-1)J(t-2)\dots J(0).
\end{equation}
We now define the matrix $D$ as
\begin{equation}
D(t)
:=
\left\vert \frac{\partial \vec{x}(t)}{\partial \vec{x}(0)} \right\vert,
\label{eq:def_D_matrix}
\end{equation}
where $\vert\dots\vert$ is the matrix (element by element) absolute value.
Since $\vert A B\vert\leq \vert A \vert \vert B \vert$, we get,
\begin{equation}
\label{eq:D}
D(t)\leq
%\vert J(t-1) \vert J(t-1) \vert \vert J(t-2) \vert \dots \vert J(0) \vert.
\vert J(t-1) \vert\,  \vert J(t-2) \vert \dots \vert J(0) \vert.
\end{equation}
Now, using Eq.~(\ref{eq:bound_cml}), we can bound,
\begin{equation}
\label{eq:bound_J}
\vert J(s) \vert\leq M A,
\end{equation}
where $[A]_{ij}=(1-\gamma)\delta_{ij}+\frac{\gamma}{2}(\delta_{i-1,j}+\delta_{i+1,j})$ is the
%conectivity
connectivity
matrix.
Finally, recalling the definition of $D(t)$ from Eq.~(\ref{eq:def_D_matrix})
(see also Eq.~(\ref{15n})), and using Eq.~(\ref{eq:D}) and~(\ref{eq:bound_J}),
we arrive at a bound for $D_{ij}$,
%\PRR{I think that also (15) and (27) must be utilized and should be mentioned}
\begin{equation}
\label{eq:bound_d_cml}
D_{ij}(t)\leq M^t [A^t]_{ij}.
\end{equation}
%Since
Observing that
$A$ is a stochastic matrix, i.e., $A_{ij}\geq 0$ and $\sum_j A_{ij}=1$,
%that implies
we can conclude
that $A^t$ is also a stochastic matrix, which ensures that $\sum_j [A^t]_{ij}=1$.
%This allow us to arrive at
Together with (\ref{eq:bound_d_cml}) we thus obtain
\begin{equation}
\sum_j D_{ij}(t)\leq M^t,
\end{equation}
which confirms that this simple coupled map lattice fulfills our main
%hypothesis in Eq.~
requirement
(\ref{eq:dyn_hyp}).

It is instructive to note that $A^t$ is bounded by the heat kernel~\cite{barlow17_random_walks_heat_kernel},
\begin{equation}
[A^t]_{ij}\leq \frac{C}{\sqrt{t}}\exp\left[ -\frac{\vert i-j \vert^2}{c t} \right],
\end{equation}
where $c$ and $C$ are constants that depend only on $\gamma$. This induces a bound over $D_{ij}$ through Eq.~(\ref{eq:bound_d_cml}) which clearly exhibit the interplay between temporal and spatial decays of the derivatives $\left\vert \frac{\partial x_j(t)}{\partial x_i(0)} \right\vert$, as discussed in the previous section.

Thus, for any local map function which satisfies Eq.~(\ref{eq:cond_local_map}), our result ensures that any macroscopic observable $g(x_0,x_1,\dots,x_n)$ which satisfies conditions
%in Eqs.~(\ref{eq:finite_der}), (\ref{eq:bounded_dif}) and (\ref{eq:ci_limit}),
(\ref{eq:finite_der})-(\ref{eq:ci_limit}),
will exhibit dynamical typicality, as long as the initial conditions are drawn independently for each dynamical variable $x_i$.

We will illustrate this result with an example. As local map function $f(x)$ we will choose the Bernoulli map,
\begin{equation}
\label{eq:bernoulli}
f(x) :=
\begin{cases}
    \text{$2x$} & \text{for $0\leq x<1/2$} \\
    \text{$2x-1$} & \text{for $1/2 \leq x <1$.}
\end{cases}
\end{equation}
Although this map is not strictly differentiable in all the domain, it is piece-wise differentiable, the derivative is non-existent only in a measure zero set. Moreover,
\begin{equation}
\left\vert \frac{df}{dx} \right\vert=2,
\end{equation}
almost everywhere. Then, we can safely choose $M=2$. The results in this section then ensure that the coupled map lattice defined according to Eq.~(\ref{eq:cml}) will satisfy,
\begin{equation}
\sum_j D_{ij}(t)\leq 2^t.
\end{equation}

For our example, we will use the following observable,
\begin{equation}
\label{eq:rugosity}
w^2(t) := \frac{1}{n}\sum_{i=0}^{n} \left(x_i(t)-\frac{1}{n}\sum_{j=0}^{n}x_j(t)\right)^2,
\end{equation}
which is a standard observable in the field of stochastic surface growth~\cite{katzav05_cml_diff_eqs}
%, which
that measures the rugosity of the spatial profile. Its derivative,
%is,
\begin{equation}
\frac{\partial w^2}{\partial x_i}=\frac{2}{n}\left(x_i-\frac{1}{n}\sum_j x_j\right),
\end{equation}
%which is bounded,
can be bounded as
\begin{equation}
\left\vert \frac{\partial w^2}{\partial x_i} \right\vert \leq \frac{2}{n},
\end{equation}
where we have used that $\vert X \vert=1$ is the length of the interval in which the dynamical variables take values.
This implies that we can choose,
\begin{equation}
c_i=\frac{2}{n},
\end{equation}
which implies that,
\begin{equation}
\sum_i c_i^2=n \left( \frac{2}{n} \right)^2=\frac{4}{n} \xrightarrow[n \to \infty]{} 0.
\end{equation}
In particular $\sum_i c_i^2\sim n^{-1}$ which implies that the variance of this observable should scale, at most, as $n^{-1}$ for any finite times, according to the Efron-Stein bound Eq.~(\ref{eq:es_dyn}).
%In consequence, the observable $w^2(x_0,x_1,\dots,x_n)$  falls under our definition of macroscopic observable.
In conclusion, our example $w^2(x_0,x_1,\dots,x_n)$ satisfies the conditions
(\ref{eq:finite_der})-(\ref{eq:ci_limit}) and therefore is a macroscopic observable.

\begin{figure}[!h]
\begin{center}
\includegraphics[scale=0.36]{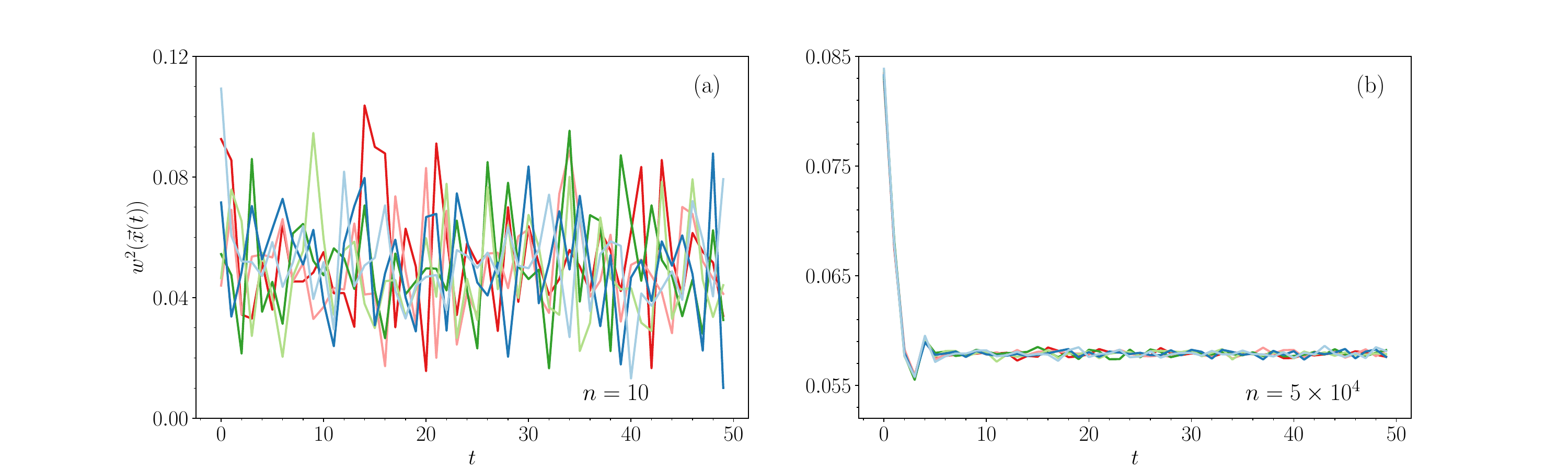}
\end{center}
\caption{
Numerically obtained results for the rugosity of the spatial profile $w^2(\vec{x}(t))$ from Eq.~(\ref{eq:rugosity}) versus time $t$
for the coupled map lattice from (\ref{eq:cml}) and (\ref{eq:bernoulli}) with $\epsilon=0.1$, and different system sizes, namely $n=10$ in (a) and $n=50000$ in (b).
The initial conditions $\vec x(0)$ have been
randomly sampled such that
each component $x_i(0)$ was uniformly and independently distributed inside
a finite interval, namely $0\leq x_i(0)<1$. The different colors correspond to
six different realizations of those random initial conditions.
}
\label{fig:fig1_cml}
\end{figure}

\begin{figure}[!h]
\begin{center}
\includegraphics[scale=0.5]{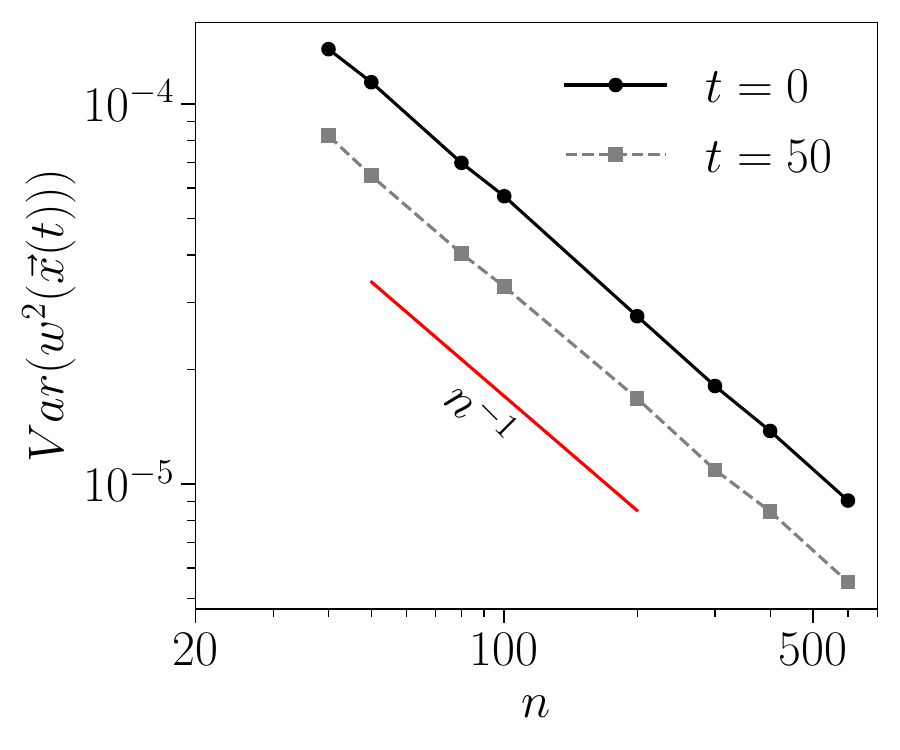}
\end{center}
\caption{
Variance of $w^2(\vec{x}(t))$ versus system size $n$ at times $t=0$ and $t=50$ for the same model as in Fig.~\ref{fig:fig1_cml}, and sampling over $5\times10^{3}$ random initial conditions.
Note the logarithmic scales. The red line serves as a guide to the eye.
%Variance of $u_t$ as a function of system size for $\alpha=2$. The variance was
%calculated sampling $5\times10^{3}$ initial conditions from an independent, uniform
%probability distribution with $0\leq x_i(0)<1$.
}
\label{fig:fig2_cml}
\end{figure}

We have performed numerical simulations
%in
of
this system to
%corroborate
illustrate
the analytical results. In Fig.~\ref{fig:fig1_cml} we show the actual trajectories for the rugosity $w^2(\vec{x}(t))$ for systems with different sizes $n$ for a small sample of initial conditions. The initial conditions for each $x_i$ are independently and uniformly sampled inside the $[0,1)$ interval. It is clearly visible that, as $n$ increases, the curves tend to collapse, which is in line with our
%results about dynamical concentration
analytical prediction of dynamical typicality
for large $n$. Moreover, in Fig.~\ref{fig:fig2_cml} we show the variance of $w^2(\vec{x}(t))$
for different times
%and verify the $n^{-1}$ scaling, as predicted by our previous arguments.
and system sizes $n$, again in agreement with the analytically predicted
$n^{-1}$ scaling behavior.

\subsection{One-dimensional
%system
systems
of coupled rotors}

We will consider a system of rotors coupled with long range interactions on a one-dimensional lattice
with open boundary consitions.
%\PRR{(Please say something why we consider long range here and short range before.)}
The Hamiltonian is
\begin{equation}
H=\sum_{i=1}^{N}\frac{p^2_i}{2}-\frac{J_N}{2}\sum_{i\neq j}\,\frac{\cos(q_i-q_j)}{\vert i-j \vert^{\alpha}},
\label{15}
\end{equation}
where the constant
\begin{equation}
J_N
:=
\frac{J}{\sup_{i}\,\sum_{j(\neq i)}\frac{1}{\vert i-j \vert^{\alpha}}},
\label{16}
\end{equation}
is introduced to ensure the extensive character of the energy~\cite{kastner14_spreading_long_range}
even for $\alpha<1$,
and where $J$ is considered as arbitrary but fixed.
We also remark that the last sum in Eq.~(\ref{15}) is meant to run over all
$i,j\in\{1,...,N\}$ with the property $i\not=j$, while the sum in
Eq.~(\ref{16}) is meant, for any given $i\in\{1,...,N\}$, to run over all
$j\in\{1,...,N\}\setminus\{i\}$.
%\PRR{I did not understand why and where these equations are needed. Maybe it would be sufficient
%(and better) to explain more precisely the meaning of the two double sums.}
%\NNR{I use the boundary conditions in the numerical simulations. Which double sum
%in particular do you think we have to clarify?}
%\PRR{I think that now everything is uniquely fixed. The fact that we work with open boundary conditions
%follows as a consequence, but we do not need any additional specifications such as for instance
%$p_{0}=q_{0}=0$ or $p_{N+1}=q_{N+1}=0$ since all these quantities actually never appear
%in the above equations. Possibly also the reply letter needs adaptation.}
%We will work with open boundary conditions, $p_{0}=q_{0}=0$ and $p_{N+1}=q_{N+1}=0$.
The dynamical variables are $\vec{x}=(\vec{q},\vec{p})$, and $n=2N$. For every rotor, the angular coordinate $q$ is defined on the bounded interval $X_q=[0,2\pi)$. However, angular momenta are defined on the entire real line $X_p=\mathbb{R}$.
This prevent us from applying the general results of the previous section without some additional conditions.
We will choose to work with distributions of initial conditions which have support on a bounded interval $\tilde{X}^{(i)}_p$
for each random momenta, which allows us to apply the results of the previous section replacing $X_p$ by $\tilde{X}^{(i)}_p$.
%Given that the system obeys Hamiltonian dynamics, this naturally induces a bound for the momenta \textit{at all times} given by $\vert p_i(t) \vert<\sqrt{2E_{\mathrm{max}}}$. In other words, for any given initial configuration with non-zero probability, all the momenta are restricted to vary inside the interval $X_p=[-\sqrt{2E_{\mathrm{max}}},\sqrt{2E_{\mathrm{max}}}]$, which ensures the applicability of the results in the previous section.
In particular, under these conditions, the energy
$H$
of the system of rotors constitutes a macroscopic observable according to our definition
%which implies
(\ref{eq:finite_der})-(\ref{eq:ci_limit}), implying
that, for large $N$, the energy tends to concentrate sharply around a typical value.
This result follows from the fact that the kinetic term in the Hamiltonian
(first sum on the right-hand side of (\ref{15}))
is a sum of independent random variables.
%On the other hand,
Moreover,
it is possible to show (see calculations below) that the interaction term (second sum on the right-hand side of (\ref{15}))
is also a macroscopic observable.
%\PRR{It seems that $E_{max}$ should grow with $n$, while the $X_i$ were silently assumed to be independent of $n$.
%What is the problem to work with finite $X_p$ in the first place? This may still amount to very reasonable
%non-equilibrium initial conditions. For instance some Gaussian distributions, independent of $n$,
%and possibly with non-zero mean, whose tails are cut off.} \NNR{You are right, all results apply by replacing the natural domain $X_p=\mathbb{R}$ by the finite support of the probability distribution $\tilde{X}^{(i)}_p$. There is no need to bound the individual momenta with the energy. I have modified the paragraph accordingly.}
%\PRR{This statement does not seem to be immediately obvious! I guess one needs to use
%the calculations which are done later on. It may also be useful to say that ``energy'' refers to
%the Hamiltonian.} \NNR{You are right.}

With the basic setup in place, we proceed to check
%the hypothesis
our main condition  (\ref{eq:dyn_hyp})
on the dynamics.
In Ref.~\cite{kastner14_spreading_long_range}, M\'{e}tivier et al. obtained the following bounds for the derivatives of our interest,
\begin{eqnarray}
\label{eq:bounds_rotors_first} D^{qq}_{ij}(t)
&:=&
\left| \frac{\partial q_j(t)}{\partial q_i(0)}  \right|\leq \frac{\cosh (vt)-1}{\vert i-j \vert^{\alpha}},\\
D^{qp}_{ij}(t)
&:=&
\left| \frac{\partial q_j(t)}{\partial p_i(0)}  \right|\leq \frac{\sinh \vert vt\vert-\vert vt\vert}{v\vert i-j \vert^{\alpha}},\\
D^{pq}_{ij}(t)
&:=&
\left| \frac{\partial p_j(t)}{\partial q_i(0)}  \right|\leq \frac{v\sinh \vert vt\vert}{\vert i-j \vert^{\alpha}},\\
\label{eq:bounds_rotors_last} D^{pp}_{ij}(t)
&:=&
\left| \frac{\partial p_j(t)}{\partial p_i(0)}  \right|\leq \frac{\cosh (vt)-1}{\vert i-j \vert^{\alpha}},
\end{eqnarray}
where $v$ is a constant with a finite $N\rightarrow\infty$ limit.
%We
%verify
%\PR{have verified}
%\PRR{(numerically or analytically?)}
%that all derivatives grow exponentially for large $t$ \NNR{This phrase is wrong, what increases exponentially are the bounds on the functions, not the functions themselves, we can erase this sentence.}.
Moreover, for $\alpha>1$
\begin{equation}
\label{eq:zeta}
\lim_{n\rightarrow\infty}\sum_{j(\neq i)}\frac{1}{\vert i-j \vert^{\alpha}}=2\zeta(\alpha),
\end{equation}
where $\zeta(\alpha)$ is the Riemann zeta function.
%, which implies that the
It follows that the
sums $\sum_{j}D^{ab}_{ij}(t)$ are all convergent for large $N$,
and hence the condition  (\ref{eq:dyn_hyp}) is fulfilled.
%Then, our bound ensures that any function
According to our general analytical prediction of dynamical typicality
from the previous section we can infer that any observable $g$
which verifies the conditions
%Eqs.~(\ref{eq:finite_der}), (\ref{eq:bounded_dif}) and (\ref{eq:ci_limit})
(\ref{eq:finite_der})-(\ref{eq:ci_limit})
will exhibit sharp concentration for large enough $N$ at any
%finite time
given time-point
$t\geq 0$.

As an example for such an observable $g$, we will take the potential energy density
(last term on the right-hand side of (\ref{15}))
\begin{equation}
u
:=
-\frac{J_N}{2N}\sum_{i\neq j}\,\frac{\cos(q_i-q_j)}{\vert i-j \vert^{\alpha}}.
\label{22}
\end{equation}
Since the function depends only on the angle coordinates $q_i$, we only need to bound the partial derivatives
(see Eq.~(\ref{eq:finite_der}))
\begin{equation}
\frac{\partial u}{\partial q_k}=\frac{J_N}{N}\,\sum_{j(\neq k)}\,\frac{\sin(q_k-q_j)}{\vert k-j \vert^{\alpha}}.
\end{equation}
Recalling the definition of $J_N$
in (\ref{16}) yields
\begin{equation}
\left| \frac{\partial u}{\partial q_k} \right|\leq \frac{J_N}{N}\,\sum_{j(\neq k)}\,\frac{1}{\vert k-j \vert^{\alpha}}\leq \frac{J}{N},
\end{equation}
which implies that we can chose $L=\frac{J}{N}$ (see Eq.~(\ref{eq:finite_der})). Now we are going to bound the quantity $c_i(t)=L\vert X_i \vert K_i(t)$ (see Eq.~(\ref{eq:c_dyn})). In the particular case of the potential energy
$u$ from Eq.~(\ref{22}),
we only need to bound $\sum_{j}D^{qq}_{ij}(t)<K_i(t)$
(see Eq.~(\ref{eq:dyn_hyp})), since the observable
$u$
only
%contains coordinate variables.
depends on the angle coordinates $q_i$.
Using the bounds in Eq.~(\ref{eq:bounds_rotors_first}) and Eq.~(\ref{eq:zeta}) we find that
%a suitable election would be
we can choose $K_i(t)= 2\zeta(\alpha)\,\vert \cosh(vt)-1 \vert$. Since $\vert X_i\vert=2\pi$ in our case, putting everything together, we find
%\PRR{it would be helpful to name the equations which are actually put together} \NNR{Yes, I think it's done with the previous lines.}
\begin{equation}
c_i(t)\leq\frac{4\pi J}{N}\,\zeta(\alpha)\,\vert \cosh (vt)-1 \vert.
\end{equation}
%this
This
implies that $\sum_{i}c^2_i(t)$ scales, at most, as $N^{-1}$. In other words, the potential energy density $u$ verifies all the conditions (\ref{eq:finite_der})-(\ref{eq:ci_limit})
of a macroscopic observable and therefore will show sharp concentration
for large $N$, for any $t\geq 0$.
In particular, the variance $Var(u_t)$ will scale, at most, as $N^{-1}$ for any finite time.
Note that our observable $u$ from (\ref{22}) is not a sum of independent random variables, hence all these findings
are \textit{not} simply a consequence of the central limit theorem.

\begin{figure}[!h]
\begin{center}
\includegraphics[scale=0.36]{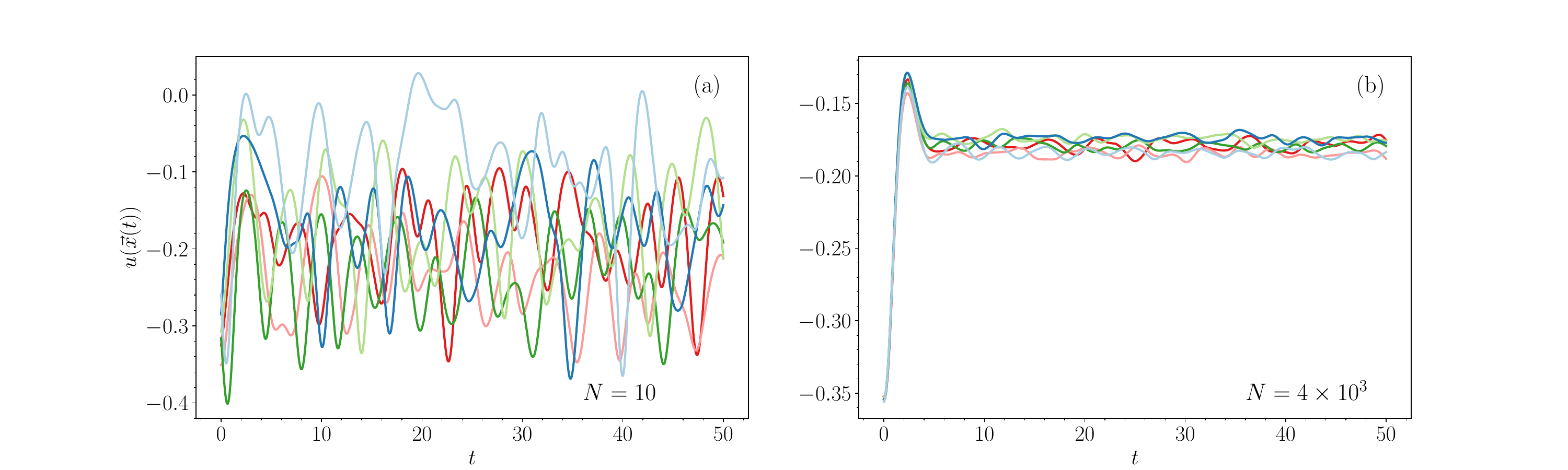}
\end{center}
\caption{
Numerically obtained results for the potential energy density $u(\vec{x}(t))$ from (\ref{22}) (see also Eq.~(\ref{6})) versus time $t$
for the model of coupled rotors from (\ref{15}) and (\ref{16}) with $J=1$, $\alpha=2$
and different system sizes, namely $N=10$ in (a) and $N=4000$ in (b).
The initial conditions $\vec x(0)=(\vec{q}(0), \vec{p}(0))$ (see also below Eq.~(\ref{16})) have been
randomly sampled such that
each component $p_i(0)$ and $q_i(0)$ was uniformly and independently distributed inside
a finite interval, namely $-1.5\leq p_i(0)<1.5$ and $-1\leq q_i(0)<1$. The different colors correspond to
six different realizations of those random initial conditions.
%Dynamics of the potential energy density starting from random initial conditions for different system sizes. Initial conditions for each $p_i(0)$ and $q_i(0)$ are uniformly and independently distributed inside a finite interval, $-1.5\leq p_i(0)<1.5$ and $-1\leq q_i(0)<1$.
%\PRR{Maybe better in color? Maybe a different choice of the intervals would result in a clearer manifestation of non-equilibrium at $t=0$?
%For instance $0\leq p_i(0)<0.1$ and $1\leq q_i(0)<2$.} \NNR{I've enlarged the intervals, the out-of-equilibrium dynamics is more evident now.}
%\PRR{Maybe repace label $u$ by $u(\vec{x}(t))$.}
}
\label{fig:fig1}
\end{figure}

\begin{figure}[!h]
\begin{center}
\includegraphics[scale=0.5]{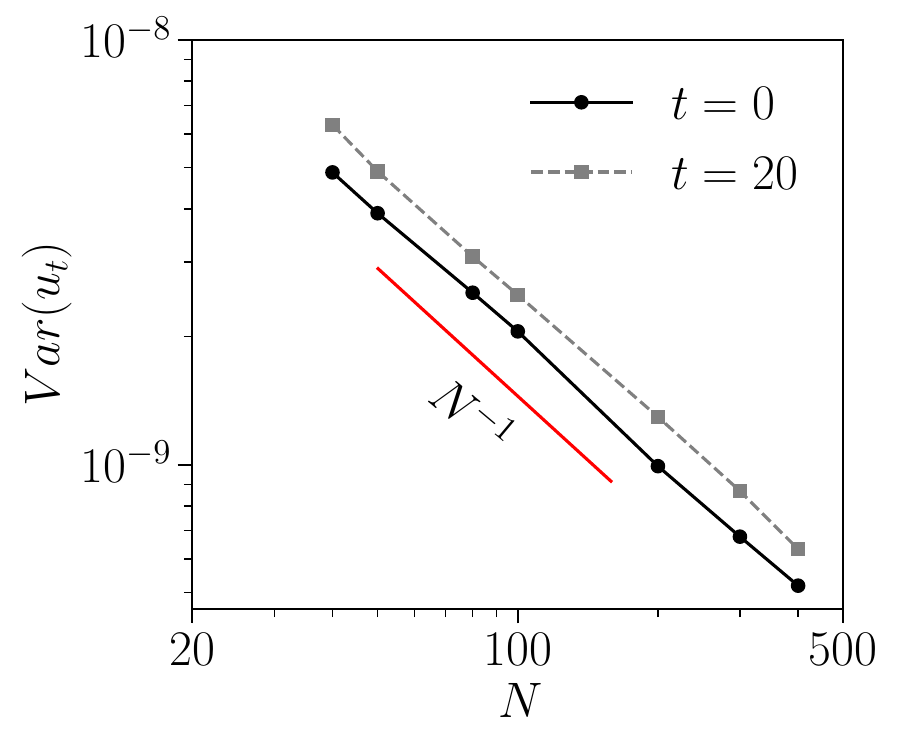}
\end{center}
\caption{
Variance of $u(\vec{x}(t))$ versus system size $N$ at times $t=0$ and $t=20$ for the same model as in Fig.~\ref{fig:fig1}
but now
with
%$\alpha=2$,
$0\leq p_i(0)<0.1$,
$0\leq q_i(0)<0.1$, and sampling over $5\times10^{3}$ random initial conditions.
Note the logarithmic scales. The red line serves as a guide to the eye.
%Variance of $u_t$ as a function of system size for $\alpha=2$. The variance was calculated sampling $5\times10^{3}$ initial conditions from an independent, uniform probability distribution with $0\leq p_i(0)<0.1$ and $0\leq q_i(0)<0.1$.
}
\label{fig:fig2}
\end{figure}

We have performed numerical simulations to
%test
quantitatively illustrate
the predictions of our analytical results. In Fig.~\ref{fig:fig1} we show the actual trajectories for the potential energy density for systems with different sizes $N$ for a small sample of initial conditions. The initial conditions for each momentum and coordinate are independently and uniformly sampled inside a given, finite, interval. It is clearly visible that, as $N$ increases, the curves tend to collapse, which is in line with our results about dynamical concentration for large $N$. Moreover, in Fig.~\ref{fig:fig2} we show the variance of $u_t$ for different times and verify the $N^{-1}$ scaling, as predicted by our previous arguments.

Additionally, we
have checked
(not shown)
that $Var(u_t)$ remains bounded and close to its initial value at large times even for small system sizes ($N\sim 10$), which implies that our bounds are not tight -- concentration of measure can be verified in conditions where our bounds are uninformative. Moreover, we have checked that dynamical typicality is present even for $\alpha<1$, a value of $\alpha$ for which the bounds in Eqs.~(\ref{eq:bounds_rotors_first})-(\ref{eq:bounds_rotors_last}) do not ensure that $\sum_{j}D^{ab}_{ij}(t)$ is convergent for large $N$.

\section{Summary and Conclusions}
\label{sec:conclusions}

The main achievement of our present study is a general analytical theory
which predicts the occurrence of dynamical typicality for a wide variety of
high-dimensional classical models, both in continuous and discrete
time, and covering Hamiltonian as well as dissipative systems.
The first key prerequisite for the applicability of our theory is that the
initial conditions are randomly sampled according to a probability
distribution in which the dynamical variables are statistically independent.
Second, we only admit macroscopic observables in the sense that
they must satisfy the conditions (\ref{eq:finite_der})-(\ref{eq:ci_limit}).
Essentially, this means that the observable may depend on
many or even all the systems's microscopic dynamical
variables, but depends sufficiently weakly on any single
one of them.
Our third key prerequisite is that the dynamics must satisfy the condition
(\ref{eq:dyn_hyp}), essentially meaning that the influence of a local
perturbation in the initial state decays, at any given instance of time,
sufficiently fast with distance.
We have provided general intuitive arguments as well as two
specific examples in support of our hypothesis that these
prerequisites should be satisfied by a very large class of
physically interesting classical many-body systems.

%We have shown that the dynamics of \textit{macroscopic} observables in classical lattice
%systems exhibit dynamical typicality under very general assumptions for large system sizes.
%Our proof
Technically speaking, our approach
is based on general concentration of measure
%inequalities
theorems,
valid for independent probability distributions.
%The approach that we have taken is
Essentially, our main new idea was
to show that the above mentioned macroscopic observables,
when compounded with the dynamical evolution, still fall under
the conditions of
%the
these
concentration theorems.

Our dynamical typicality
%phenomenon suggests
results imply
that the dynamics of macroscopic observables is largely
independent of the microscopic details of the initial conditions.
Projector operator techniques may prove useful in the study of this
macroscopic dynamics~\cite{grabert82_pot_non-eq}.

We want to
%remarks,
remark
that, although we expect that our main condition on the dynamics, Eq.~(\ref{eq:dyn_hyp}),
% to
will
be fulfilled in a great variety of physically interesting systems, it is difficult to make general statements about its validity. As an interesting example, consider a dissipative system which posses two different attractors. Moreover, let us assume that both attractors are macroscopically different, i.e., there are macroscopic observables that take different values in the two attractors.
Then, any finite measure neighborhood around the boundary between the basins of
attraction of the two attractors constitutes a finite measure set of initial conditions
such that microscopic modifications in the states lead to macroscopically different states for sufficiently
long times.
%If the basins of attraction of the two attractors share a boundary with non-zero measure
%\PRR{I think the boundary is usually of zero measure (subset of co-dimension one)},
%then there is a set of initial conditions with non-zero measure
%\PRR{this should still remain true, since an entire neighborhood of the boundary is admitted}
%such that microscopic modifications in the states lead to macroscopically different states for sufficiently long times.
Then, if the time needed to arrive to the asymptotic regime dominated by the attractors is finite, condition (\ref{eq:dyn_hyp}) will be violated, because a microscopic difference in the initial conditions would lead to a macroscopic difference for finite times. A similar scenario may be present for Hamiltonian systems which exhibit phase transitions, with the ergodic components playing a similar role to the basins of attraction
%of
in
the previous example. These two scenarios in which our condition on the dynamics Eq.~(\ref{eq:dyn_hyp}) is not fulfilled constitute an interesting direction for future work.
%\PRR{I think in this case it is better to consider directly the left-hand side of (14) instead of the right-hand side:
%It seems to me quite obvious that there will always (or generically)
%be some $\vec x(0)$ near to the basin boundary and
%some $\Delta x_i$ such that the displaced point $\vec x(0)+\Delta x_i \vec e_i$ is on the opposite
%side of the basin boundary, and hence the left-hand side of (14) is not small for some finite $t$.
%Moreover, since we do not require that $\Delta x_i$ must be small, the set of such $\vec x(0)$
%will not be of small measure. We should first agree on what we want to say before we write it up.}
%\NNR{I think the argument using the LHS of (14) can be useful. We need to argue that in these
%cases (dissipative system with two macroscopically different attractors and Hamiltonian system
%with a phase transition) the LHS of (14) is not small in such a way that Eqs. (17) and (20) are
%violated. This is the way that seems natural to me, but there may be other valid reasoning paths.}

Finally,
our above specified requirement regarding the probability distribution of the
initial conditions may still appear quite restrictive.
However,
we would like to mention that our results may still be valid in the presence of weak correlations in the initial state, since the general concentration theorems used in our proof can be generalized in such cases~\cite{esposito24_concentration_non-indep}.

\section*{Acknowledgements}
This work was supported by the
Deutsche Forschungsgemeinschaft (DFG, German Research Foundation)
under Grant No. 355031190 within the Research Unit FOR 2692
and under Grant No. 502254252.

\newpage

\bibliographystyle{phaip}

\end{document}